\documentclass{article}     
\usepackage{latexsym}       

\def\keywords#1{\\ {\bf Key words:\ }#1}

\begin{document}
\begin{titlepage}
\title{GENERALIZED FIERZ IDENTITIES AND THE SUPERSELECTION RULE FOR
 GEOMETRIC MULTISPINORS
\thanks{Appears in: {\it Spinors, Twistors,
Clifford Algebras and Quantum Deformations (Proceedings of the
Second Max Born Symposium, Wroclaw, Poland, Sept. 24-27, 1992)} Z.
Oziewicz et al. (eds.), Kluwer Academic Publ. (1993), pp.\ 91-96;
preprint version gr-qc/9211018.  Based on conference presentation,
slides available at
http://www.clifford.org/wpezzag/talk/92wroclaw/ }}
\author{William M. Pezzaglia Jr.\thanks{Current address:
Department of Physics, Santa Clara University, Santa Clara,
CA 95053, USA, wpezzag@clifford.org} \\ \\
Department of Physics and Astronomy\\San Francisco State
University\\1600 Holloway Avenue\\San Francisco, California
94132\\U.S.A.}
\date{Version 3.0, (December 10, 2001)}

\maketitle

\begin{abstract}
The inverse problem, to reconstruct the general multivector wave
function from
 the observable quadratic densities, is solved for 3D geometric
algebra.  It is found that operators which are applied to the
right side of the wave function must be considered, and the
standard Fierz identities do not necessarily hold except in
restricted situations, corresponding to the spin-isospin
superselection rule.  The Greider idempotent and Hestenes
quaterionic spinors are included as extreme cases of a single
superselection parameter. \keywords{fierz -- multivector --
superselection -- spinors}

\end{abstract}

\end{titlepage}

\section{Introduction}

\quad In a recent paper Crawford[1] explored the inverse problem
of Dirac bispinor algebra, to reconstruct the wave function from
the observable quadratic densities.  Other authors$^{2,3}$ have
presented parallel developments for multivector quantum mechanics
in which column spinors are replaced by Clifford algebra
aggregates.  However, these expositions have only considered
restricted cases (e.g. minimal ideals) for which the multivector
analogies of the observable bispinor densities obey the standard
Fierz$^{1,2,4}$ identities.

Previously we have proposed$^{5,6}$ a more general multivector
wave function in which all the geometric degrees of freedom are
used.  To obtain the complete set of observable multispinor
densities one needs to augment the standard {\it sinistral}$^{\
6}$ operators (applied to the left side of the wave function) with
new {\it dextral}$^{\ 6}$ (right-side applied) operators, and also
{\it bilateral}$^{\ 6}$ operators (multivectors applied on both
sides of the wave function).  It is found that if and only if the
multivector wave function is restricted will the multispinor
densities obey the standard Fierz identities.  In this paper we
propose to solve the inverse problem for the general unrestricted
multivector wave function of the 8-element 3D geometric algebra
${\bf C}(2)$, i.e. the Pauli algebra.

\section{The Algebra of Standard Pauli Spinor Densities}
\quad In non-relativistic quantum mechanics, the electron is
represented by a two-component Pauli spinor.  The endomorphism
algebra (module structure on spinors) is ${\bf C}(2)$, i.e. two
by two complex matrices.  This Clifford algebra has as its basis
the 8 element group generated by 3 mutually anticommuting basis
vectors, $\{\sigma_j,\sigma_k\}=2\delta_{jk}$ for (j,k = 1,2,3),
and where $i = \sigma_1\sigma_2\sigma_3$.  As operators, their
``bilinear expectation values'' yield real densities which are
interpreted to be the projection of the spin along the j-th spatial
axis, $S_j=<\psi|\sigma_j|\psi>$.

The 4 densities $\{\rho,S_i\}$, where $\rho=<\psi | \psi>$, are
invariant with respect to the phase parameter of the spinor.  Hence
they satisfy a single constraint equation which can be derived by
substituting the projection operator into the square of the
normalization.  The magnitude of the spin is found to be
constrained by the Fierz$^{1,2,4}$ identity,
$$|{\bf S}|^2 = S^kS_k =  \rho^2  . \eqno(1)$$
The spinor can be reconstructed in terms of a $U(2)$ unitary
rotation matrix,
$$U(\lambda,\theta,\phi,\alpha)= \exp(i\alpha/2)
\exp(i\sigma_3\phi/2) \exp(i\sigma_2 \theta/2)
\exp(i\sigma_3 \lambda/2) ,   \eqno(2)$$
where $(\theta, \phi)$ are the orientation angles of the spin and
$(\lambda,\alpha)$ do not contribute in the bilinear
form $\sigma^j S_j (\theta,\phi) = \rho U \sigma_3 U^{\dagger}$.
Choosing a starting spinor to be the ``plus'' eigenstate of
$\sigma_3$, the wave function can be expressed,
$\psi(\rho,\theta,\phi,\beta)=\sqrt{\rho} U |\eta>$, where the net
unobservable phase is $\beta=\lambda+\alpha$.

\section{The Algebra of Geometric Multispinors}

\quad We consider an unrestricted multivector wave function$^5$
which has
 the same {\it 8 degrees of freedom} as the Clifford group,
$$\psi=\left(\matrix{a &c \cr b &d\cr}\right)   = (a + b \sigma_1)
{(1 + \sigma_3)\over 2} + (c + d \sigma_1){(1 +
 \sigma_3)\over 2} \sigma_1 ,   \eqno(3)$$
where $\{a,b,c,d\}$ are complex coefficients.  Note each {\it
column} of the matrix is a {\it minimal left ideal} of the algebra
and will hence behave like a column spinor$^8$ for all
{\it sinistral}\ $^6$ (left-sided ) operations.  Each {\it row}
of the matrix is a {\it minimal right ideal} of the algebra and
will behave like a row {\it isospinor} for all dextral$^6$
(right-sided) operations.  Hence the complete solution can be
interpreted$^6$ as an isospin doublet (of spinors) coupled by
now-allowable dextral application of the Pauli operators.

\subsection{Multivector Densities}

\quad A complete set of 16 generalized quadratic forms are
defined in terms of the matrix trace (i.e. half the scalar
part of the Clifford multivector)$^9$,
$$\rho=Tr(\psi^{\dagger}\psi)=Tr(\psi\psi^{\dagger})=
|a|^2+|b|^2+|c|^2+|d|^2, \eqno(4a)$$
$$S_j=Tr(\psi^{\dagger}\sigma_j\psi)=
Tr(\psi\psi^{\dagger}\sigma_j), \quad (j=1,2,3), \eqno(4b)$$
$$T_j = Tr(\psi \sigma_j \psi^{\dagger}) =
Tr( \psi^{\dagger}\psi \sigma_j) ,
\quad(j = 1,2,3) ,  \eqno(4c)$$
$$R_{jk} = Tr(\psi^{\dagger} \sigma_j \psi \sigma_k) =
Tr( \psi \sigma_k
 \psi^{\dagger} \sigma_j) ,   \quad(j,k = 1,2,3).   \eqno(4d)$$
They are interpreted to be the probability, spin, isospin
and {\it bilateral} densities respectively.  From these we can
construct the multivector densities,
$$\psi \psi^{\dagger}=(\rho+\sigma_kS^k)/2, \eqno(5a)$$
$$\psi^{\dagger} \psi=(\rho+\sigma_kT^k)/2, \eqno(5b)$$
$$\psi^{\dagger}\sigma_j \psi=(S_j+R_{jk}\sigma^k)/2, \eqno(5c)$$
$$\psi\sigma_k \psi^{\dagger}=(T_k+\sigma^jR_{jk})/2. \eqno(5d)$$

\subsection{Generalized Fierz Identities}

\quad The 16 densities are all independent of the phase parameter,
hence must satisfy 9 constraint equations.  In general these
identities have the form,
$$Tr[(\psi^{\dagger}\sigma_{\alpha} \psi)\sigma_{\beta}
 (\psi^{\dagger}\sigma_{\gamma} \psi) \sigma_{\delta} ] =
Tr[(\psi\sigma_{\beta} \psi^{\dagger})\sigma_{\gamma}
 (\psi\sigma_{\delta} \psi^{\dagger}) \sigma_{\alpha} ] \eqno(6a)$$
where the indices can take on values 0 through 3, and $\sigma_0
=1$.  The parenthesis indicate where one inserts eqs. (5abcd).
It can be shown from these relations that the bilateral
density eq.\ (4d) contains all the other densities.  Further,
we find that the magnitudes of the spin and isospin
are equal, but that eq.\ (1) is no longer valid,
$$|{\bf S}|^2 = |{\bf T}|^2 \leq \rho^2 .   \eqno(6b)$$

\subsection{Interpretation of the Bilateral Density}

\quad Counting degrees of freedom,we see that there is one free
internal ``hidden variable'' contained in $R_{jk}$ which does
not affect the other densities.  To gain some insight as to the
nature of this parameter we consider the class of unitary
(hence $\rho$ invariant) transformations that will leave the
densities $\{S^j, T^j \}$ invariant, but modify the bilateral
density.

The special case sinistral operator, $U(\lambda) =
 \exp[i\sigma_j S^j/(2 |{\bf S}|)]$, will leave the spin
invariant (as well as the isospin) as it corresponds to a
rotation about the spin axis by angle $\lambda$.  The bilateral
density will be modified by this transformation, hence we should
be able to parametrize $R_{jk}$ in terms of the densities
$\{ \rho, S^j, T^j \}$ and a bilateral phase angle $\lambda$.

\section{Inverse Theorem and Superselection Rule}

\quad We assert that the projection operator for the multivector
wave function has the bilateral form,
$\psi= ( \rho \psi +  S_k \sigma^k \psi +  \psi \sigma_k T^k +
 \sigma^j \psi \sigma^k R_{jk} )/(4\rho)$.

\subsection{Inverse Theorem}

\quad The multivector wave function can be reconstructed from
the observable densities by a applying the projection operator
to an arbitrary starting solution $\eta$, and renormalizing.
Hence we assert,
$$\Psi(\alpha,S^k,T^k,R^{jk})= e^{i\alpha} ( \rho\eta +
S^k \sigma_k \eta +  \eta \sigma_k T^k +  \sigma^j \eta
\sigma^k R_{jk} )/N  , \eqno(7a)$$
where $\alpha$ is a phase factor and $\eta$ is an arbitrary
starting multivector subject only to the normalized trace constraint
$Tr(\eta^{\dagger}\eta) = 1$.

The normalization factor is most directly determined by requiring
the reconstructed wave function to reproduce the probability
density eq.\ (4a),
$${N^2}=4[\rho + Tr(\eta^{\dagger} \sigma^j\eta\sigma^k R_{jk})]+
 2 [Tr(\eta^{\dagger} \eta \sigma^k T_k) +
 Tr(\eta \eta^{\dagger} \sigma^k S_k) ] ,   \eqno(7b)$$
where identities have been used to reduce the quadradic terms to
linear ones in terms of the observable densities. This construction
will fail if eq.\ (7b) yields zero, in which case a different
starting solution should be used.

\subsection{Special Classes of Solutions}

\quad In order to insure a scalar norm, Hestenes[3,9] proposed a
unitary or {\it quaternionic} solution which has 5 parameters,
$$\psi(\alpha,\rho,\lambda,\theta,\phi)= \sqrt{\rho\over 2} \
 U(\lambda,\theta,\phi,\alpha)=\sqrt{\rho\over 2}\ e^{i\alpha/2}\
[ r +  i\sigma^j  B_j ]  ,  \eqno(8)$$ where unitary matrix
$U(\lambda,\theta,\phi,\alpha)$ is given by eq. (2).  The
alternate quaternionic {\it Cayley-Klein} components $\{r, B_j\}$
are all real numbers, subject to constraint $r^2 + B^2 =1$.
Only 4 parameters are however needed to describe an electron,
hence Hestenes (arbitrarily?) sets the parameter $\alpha$ to zero.

This unitary class of solutions is synonymous with {\it zero}
magnitude spin and isospin as defined by eqs.\ (4bc).  The bilateral
density eq.\ (4d) is proportional (by a factor of $\rho$) to the
$O(3)$ rotation matrix $R(\lambda,\theta,\phi)$ associated with
the $U(2)$ matrix $U(\lambda,\theta,\phi,\alpha)$.  This allows
Hestenes to make an alternate definition of a ``spin'' vector in
terms of the bilateral density,
$S'_j = R_{j3} = Tr(\psi^{\dagger}\sigma_j\psi \sigma_3) =
{1 \over 2}Tr( U\sigma_3 U^{\dagger}\sigma_j) \rho$.  It is easily
verified that $R_{jk}$ is invariant with respect to the $\lambda$
parameter of the unitary matrix, allowing Hestenes to reinterpret
it as quantum phase, and dextrally applied $i\sigma^3$ as the
quantum phase generator (replacing the usual commuting $i$).

In contrast, Greider[7] proposed an idempotent spinor which has
the algebraic form of the projection operator,
$$\psi= e^{i\alpha} {(\rho + S_k \sigma^k)\over 2\sqrt{\rho}}=
\sqrt{\rho}\ U(0,\theta,\phi,\alpha) {(1 + \sigma^3) \over 2}
U^{\dagger}(0,\theta,\phi,-\alpha) ,    \eqno(9)$$
where the magnitude of the spin is subject to the standard
Fierz constraint of eq.\ (1).  This makes the determinant zero,
hence the wave function is of the ``singular class'' distinctly
different from the ``unitary class'' discussed above.  There are
only 4 degrees of freedom, exactly that needed to describe a single
Pauli spinor (i.e. isospin is everywhere parallel to spin).

Isospin degrees of freedom can be re-introduced by applying a
dextrad rotation operator to eq.\ (10).  Equivalently, consider
the following factorized idempotent form,
$$\psi = \sqrt{\rho}\enskip U(\lambda,\theta_S,\phi_S,\alpha)
\enskip {(1 + \sigma^3) \over 2} \enskip
U^{\dagger}(\lambda,\theta_T,\phi_T,-\alpha) ,  \eqno(10a)$$
$$= e^{i\alpha}\  (\rho + S_k \sigma^k) (\rho + T_j \sigma^j)
\  [4\rho( \rho^2 + S_k T^k )]^{-{1 \over 2}} ,  \eqno(10b)$$
where the singularity constraint eq.\ (1) still holds.  The angles
$\{\theta_S,\phi_S\}$ give the orientation of the spin, while
$\{\theta_T,\phi_T\}$ that of the isospin.  Our solution has 6
degrees of freedom, exactly that which is needed to describe an
isospin doublet of Pauli spinors (i.e. two particle generations
in the family).  The net phase $\beta=\lambda + \alpha$ shows
$\lambda$ is indistinguishable from parameter $\alpha$, hence
$R_{jk} =\rho S_j T_k/|{\bf S}|^2$, has no $\lambda$ dependence.

\subsection{Superselection Parameter}

\quad Our multispinor solution is subject to a {\it spin-isospin
superselection rule}$^{10}$.  While certain linear combinations are
allowable, the superposition of ``spin \& isospin up'' with
``spin \& isospin down'' would yield a ``forbidden'' unitary class
solution with zero spin and isospin.  Equivalently such a state is
inaccessible by any spin/isospin rotation from a ``spin \& isospin
up'' state.  Mathematically this constraint manifests as requiring
the determinant of our wave function to be zero.

Consider a new superselection parameter $\delta$, defined: $| {\bf
S} | = \rho \cos\delta$,
$$\Psi(\alpha,\rho,\lambda,\delta,\theta_S,\phi_S)= \sqrt{\rho}
 \enskip e^{i\alpha} \exp(i\sigma^k n_k \lambda/2 )\enskip
{ (1 + e^{i\delta} n_k \sigma^k)/2} ,   \eqno(11)$$ where
$n_k(\theta_S,\phi_S) = S_k/|{\bf S}|$.  For $\delta=0$ the wave
function becomes a Greider[7] idempotent with zero determinant, and
when the spin vanishes in the limit of $\delta=\pi/2$, the solution
is of the Hestenes[3,9] quaternionic form.  Note the bilateral
phase $\lambda$ is independent of the ordinary imaginary phase
$\alpha$ for $\delta >0$.

The remaining two isospin degrees of freedom can be reintroduced
as before by a dextrad rotation operator.  A complete
parameterization of the general 8 degree of freedom solution can
be expressed in polar form,
$$\Psi( \rho, \alpha, \lambda, \delta, \theta_S,\phi_S
,\theta_T,\phi_T)= \sqrt{\rho}\enskip U(\lambda,\theta_S,
\phi_S,\alpha) {(1 + e^{i\delta}  \sigma_3)\over 2}
 U^{\dagger}(\lambda,\theta_T,\phi_T,-\alpha) . \eqno(12)$$

\section{Summary}

\quad We have solved the inverse problem for the completely
general eight degree of freedom wave function of 3D geometric
space.  Our results are more general than other treatments in
that a more complete set of quadratic multispinor densities is
introduced which includes sinistral, dextral and bilateral
operations.  The 16 densities satisfy generalized Fierz-type
identities.  The new {\it bilateral density} is found to contain
one new independent ``hidden'' variable which does not affect
the more familiar probability, spin and isospin densities.  It
is an open question as to whether this quantity can be
physically measured, or is unobservable like the overall quantum
phase parameter.

The standard Fierz identities (for column spinors) are found not
to hold except for a restricted singular class of wave functions.
This appears to be a manifestation of the spin-isospin
superselection rule, and may be the critical constraint which
classifies the solution as being a fermionic particle.  A
continuous superselection parameter is introduced for which the
singular class of solutions (which includes the Greider
idempotent form) is one extreme case; the Hestenes
quaternionic spinor form is at the other extreme.

Extending the work to 4D Minkowski space with a 16 degree of
freedom wave function we will find 136 quadratic forms, which
obey 121 generalized identities.  One or more new ``hidden''
variables will be found, and the standard Fierz identities will
not be valid except for a restricted wavefunction, corresponding
to the charge superselection rule of bispinors.

\end{document}